\documentclass[twocolappendix,appendix,numberedappendix,iop,apj]{emulateapj}
\usepackage{blindtext}
\usepackage{lipsum}
\usepackage{amssymb,amsmath,amsfonts,amstext,graphicx,aas_macros}
\usepackage{natbib}
\usepackage{color}
\usepackage{times}
\usepackage[colorlinks=true, citecolor=blue, linkcolor=black]{hyperref}
\usepackage[all]{hypcap}
\usepackage{longtable}
\usepackage{float}
\usepackage[toc,page]{appendix}
\usepackage[caption=false]{subfig}
\usepackage{mwe}

\defcitealias{vogelsberger2016}{V16}
\shortauthors{Sameie et al.}
\shorttitle{Mass Functions In Interacting DM Models}
\begin{document}
\setlength{\abovedisplayskip}{10pt}
\setlength{\belowdisplayskip}{2pt}
\title{\large The effect of dark matter-dark radiation interactions on halo abundance - a Press-Schechter approach}

\author{\normalsize Omid Sameie$^{1}$\altaffilmark{$\dagger$}, Andrew J. Benson$^{2}$, Laura V. Sales$^{1}$\altaffilmark{$*$}, Hai-bo Yu$^{1}$, Leonidas A. Moustakas$^{3}$\\
and Peter Creasey$^{1}$}
\affil{
$^{1}$ Department of Physics and Astronomy, University of California, Riverside, CA 92521 USA\\
$^{2}$ Carnegie Observatories, 813 Santa Barbara Street, Pasadena, CA 91101, USA\\
$^{3}$ Jet Propulsion Laboratory, California Institute of Technology, 4800 Oak Grove Dr, Pasadena, CA 91109, USA}
\altaffiltext{${\dagger}$}{NASA MIRO FIELDS Fellow}
\altaffiltext{$*$}{Hellman Fellow}
\email{osame001@ucr.edu}

\begin{abstract}
We study halo mass functions with the Press-Schechter formalism for interacting dark matter models, where matter power spectra are damped due to dark acoustic oscillations in the early universe. After adopting a smooth window function, we calibrate the analytical model with numerical simulations from the ``effective theory of structure formation" (ETHOS) project and fix the model parameters in the high mass regime, $M_{\rm h}\gtrsim3\times10^{10}\;{\rm M}_{\odot}$. We also perform high-resolution cosmological simulations with halo masses down to $M_{\rm h}\sim10^8\;{\rm M}_{\odot}$ to cover a wide mass range for comparison. Although the model is calibrated with ETHOS1 and CDM simulations for high halo masses at redshift $z=0$, it successfully reproduces simulations for two other ETHOS models in the low mass regime at low and high redshifts. As an application, we compare the cumulative number density of haloes to that of observed galaxies at $z=6$, and find the interacting dark matter models with a kinetic decoupling temperature below $0.5\ \rm{keV}$ is disfavored. We also perform the abundance-matching analysis and derive the stellar-halo mass relation for these models at $z=4$. Suppression in halo abundance leads to less massive haloes that host observed galaxies in the stellar mass range $M_*\simeq 10^5\textup{-}10^7\ {\rm M}_{\odot}$. 
\end{abstract}
\maketitle
\section{Introduction}

The cold dark matter (CDM) paradigm has been extremely successful at explaining numerous astrophysical phenomena on galactic and extra-galactic scales \citep[][]{seljak2005,percival2007,vogelsberger2014b,vogelsberger2014,planck2016}. However, there have been various reports pointing toward both small-scale \citep{moore1994,flores1994,klypin1999,moore1999,boylan-kolchin2011,oman2015} and large-scale issues~\citep{macCrann2015,riess2016,addison2018}. While these anomalies on different scales could be due to either systematic observational uncertainties \citep{kitching2016,joudaki2017,kim2017} or baryonic physics \citep{pontzen2012,brook2013,santos2018,gk2017}, there has been a growing interest in work within the framework of non-CDM models to address these difficulties \citep[e.g. see][]{abazajian2017,tulin2017,buen-abad2018}.

For example, DM with non-zero free streaming velocities suppresses the matter power spectrum and delays halo formation, resulting in a lower number density of virialized structures and less concentrated DM haloes \citep{lovell2014,menci2018}. Moreover, strong DM self-interactions, through kinematic thermalization, tie the DM distributions to the baryonic ones \citep{kaplinghat2014,elbert2018,sameie2018} such that it potentially reduces the tension in some of the small-scale puzzles \citep{vogelsberger2012,zavala2013,rocha2013,peter2013,kamada2017,creasey2017,robertson2018b,robertson2018a,vogelsberger2018,valli2018,ren2018}. DM could also be coupled to dark radiation such that this extra relativistic component could potentially explain the tension in the measurements of $H_0$ from local and CMB observations, and also, through damping the power spectrum via \emph{dark acoustic oscillations} (DAO), reduce the tension in $\sigma_8$ measurements and possibly the missing satellites problem \citep[e.g. see][]{boehm2014,vogelsberger2016,chacko2016,brust2017}. This rich phenomenology of ``interacting" DM models has led several authors to categorize different DM interactions based on their astrophysical predictions \citep{boehm2002,cyr-racine2016,murgia2017}, and study the astrophysical constraints on their model parameters \citep{vogelsberger2016, lovell2017,huo2017,pan2018,diaz2018}.

A main feature of these DM models when compared to standard CDM is their predictions for the abundance of DM structures in different mass regimes. Numerical simulations, and semi-analytical modeling based on the extended Press-Schechter approach \citep{press1974,bond1991,bower1991,lacey1993} have been utilized to study halo and subhalo mass functions within the context of non-CDM scenarios \citep{benson2013,schneider2013,buckley2014,schneider2015,schneider2017}. These authors have shown that the cutoff in the linear theory power spectrum suppresses the halo mass function at low masses, and hence it is possible to test these models with observations to constrain the mass function. In practice, the semi-analytical approach needs to be calibrated with numerical simulations to make reliable prediction for mass functions. Moreover, it is well-known that the choice of window function in Press-Schechter formalism has a significant impact on the predictions of the model for the suppression of the mass function in the small mass regime where deviations from CDM are expected \citep[see e.g.][]{benson2013,leo2018}. 

In this paper, we use both the semi-analatyical method and N-body simulations to study mass functions for interacting DM models, where DM is coupled to dark radiation via a force mediator. For the sake of convenience, we mainly focus on the power spectra used in the ETHOS project \citep[][]{cyr-racine2016,vogelsberger2016}, and perform the calibration analysis by comparing the analytical predictions with the simulations in the halo mass range above $3\times10^{10}\;{\rm M}_{\odot}$ at $z=0$. We further perform cosmological simulations with improved mass resolution to test the model in the low mass regime, $10^8\textup{--}10^{10}\;{\rm M}_{\odot}$, at different redshifts. The analytical model exhibits remarkable universality, i.e., once calibrated with respect to joint data points from CDM and ETHOS1 simulations in the high mass range at $z=0$, it accurately predicts the halo mass functions for other ETHOS models in different mass regimes at higher redshifts/earlier times.

The prediction of the halo abundance at high redshifts is particularly interesting. Using the observed abundance of ultra-faint high redshift galaxies \citep{menci2017,livermore2017}, we apply our analytical model to constrain the interacting DM models and compare the results with those derived from the Lyman-$\alpha$ observations \citep{huo2017}. In addition, we use the model to study the impact of suppression in matter power spectrum on the low mass tail of the stellar-halo mass relation. We take the observed stellar mass function at $z=4$ from \citet{song2016} and perform an ``abundance matching" analysis \citep{vale2004,vale2006,moster2010,guo2010,moster2013,behroozi2013} to assign halo mass to the observed galaxies at the redshift for each of the DM models considered in this work. Our goal is to show how these non-trivial DM interactions changes DM content of galactic systems through the matching procedure.

The structure of this paper is organized as follows: In Sec.~\ref{sec:method}, we introduce interacting DM models and ingredients for constructing halo mass functions in the Press-Schechter framework, and we also discuss cosmological simulations carried out in this work. We present our main results in Sec.~\ref{sec:results} and summarize in Sec.~\ref{sec:conc}.

\section{Methodology}\label{sec:method}
We work within the framework of the Press-Schechter formalism to compute halo mass functions. We use the following cosmological parameters: $\Omega_{\rm m}=0.302$, $\Omega_{\Lambda}=0.698$, $\Omega_{\rm b}=0.046$, $h=0.69$, $\sigma_{8}=0.839$, and $n_{\rm s}=0.967$ consistent with \citet{planck2016}. Throughout this work, we define halo mass as the mass enclosed by a sphere with average density equal to the virial overdensity $\Delta_{\rm vir}(z)$ \citep{bryan1998} times the critical density. This mass definition resembles closely the redshift evolution predicted by the analytic model \citet{despali2016}. The Press-Schechter formalism requires three elements, i.e., the matter power spectrum, barrier height in the excursion set approach and the solution for the distribution of first crossing events, as we will discuss in detail later. In our analysis, we use cosmological simulations in \citet[][hereafter V16]{vogelsberger2016} to calibrate the fitting formula for distribution of first crossing in mass scales above $3\times10^{10}\ {\rm M}_{\odot}$. 

In order to resolve halo abundances on lower mass scales and at different redshifts, we run cosmological N-body simulations for the benchmark models in \citetalias{vogelsberger2016}. Note that we do not include DM-DM self-interactions in the simulations, as their effect is negligible on the abundance of the haloes. We compute the power spectra using a modified version of the Boltzmann code {\sc CAMB} \citep{lewis2002,cyr-racine2016} to include DM-dark radiation interactions, and generate the initial conditions at $z = 127$ with two different periodic box sizes $\text{L}=10\ \text{Mpc}/h$ and $20\ \text{Mpc}/h$ with the code {\sc N-GenIC} \citep{springel2001,springel2005}. Our simulations are performed using $256^3$ and $512^3$ particles yielding DM particle mass resolutions of $9.04\times 10^5\;\rm M_{\odot}$ and $7.18\times 10^6\; \rm M_{\odot}$ and spatial resolutions of $\epsilon=2.5\ \text{kpc}/h$ and $5\ \text{kpc}/h$ (Plummer-equivalent softening length). In order to compare the halo mass functions predicted in the Press-Schechter model, we use results from simulations with L=10 Mpc/$h$ and $m_{\rm p}=7.18\times 10^6\rm M_{\odot}$ (referred to as ${\rm L}_{10}$). Other simulations are used to perform convergence and resolution tests as shown in Appendix~\ref{sec:A1}. Table~\ref{table:sim} summarizes the details of our simulations. We use the code {\sc Arepo} \citep{springel2010} to run simulations. Haloes and subhaloes are identified by the friends-of-friends \citep{davis1985} and {\sc SUBFIND} \citep{springel2001} algorithms which we use to construct mass functions. 

\begin{table}
\caption{\centering Parameters for our cosmological simulations.} 
\label{table:sim}
\begin{center}
\begin{tabular}{cccc}
\hline
\text{DM model} & L (Mpc/$h$) & $N_{\rm p}$ & $m_{\rm p} (\rm M_\odot)$\\
\hline
\hline
\text{ETHOS1} & $10$ & $256^3$ & $7.18\times10^6$ \\
\text{ETHOS1} & $10$ & $512^3$ & $9.04\times10^5$ \\
\text{ETHOS2} & $10$ & $256^3$ & $7.18\times10^6$ \\
\text{ETHOS3} & $10$ & $256^3$ & $7.18\times10^6$ \\
\text{ETHOS3} & $20$ & $512^3$ & $7.18\times10^6$ \\
\hline
\end{tabular}
\end{center}
\raggedright {\bf Note.} The second column (L) is the simulation box size in ${\rm Mpc}/h$, the third column ($N_{\rm p}$) is the total number of particles, and the last column ($m_{\rm p}$) is the mass resolution in $\rm M_{\odot}$. Our simulations take matter power spectra of the ethos models, but do not include DM self-interactions that have negligible effects on the halo mass functions for the mass scales that we are interested.
\end{table}

\subsection{Power spectrum}
\label{sec:power}

\begin{figure}
\includegraphics[width=\columnwidth]{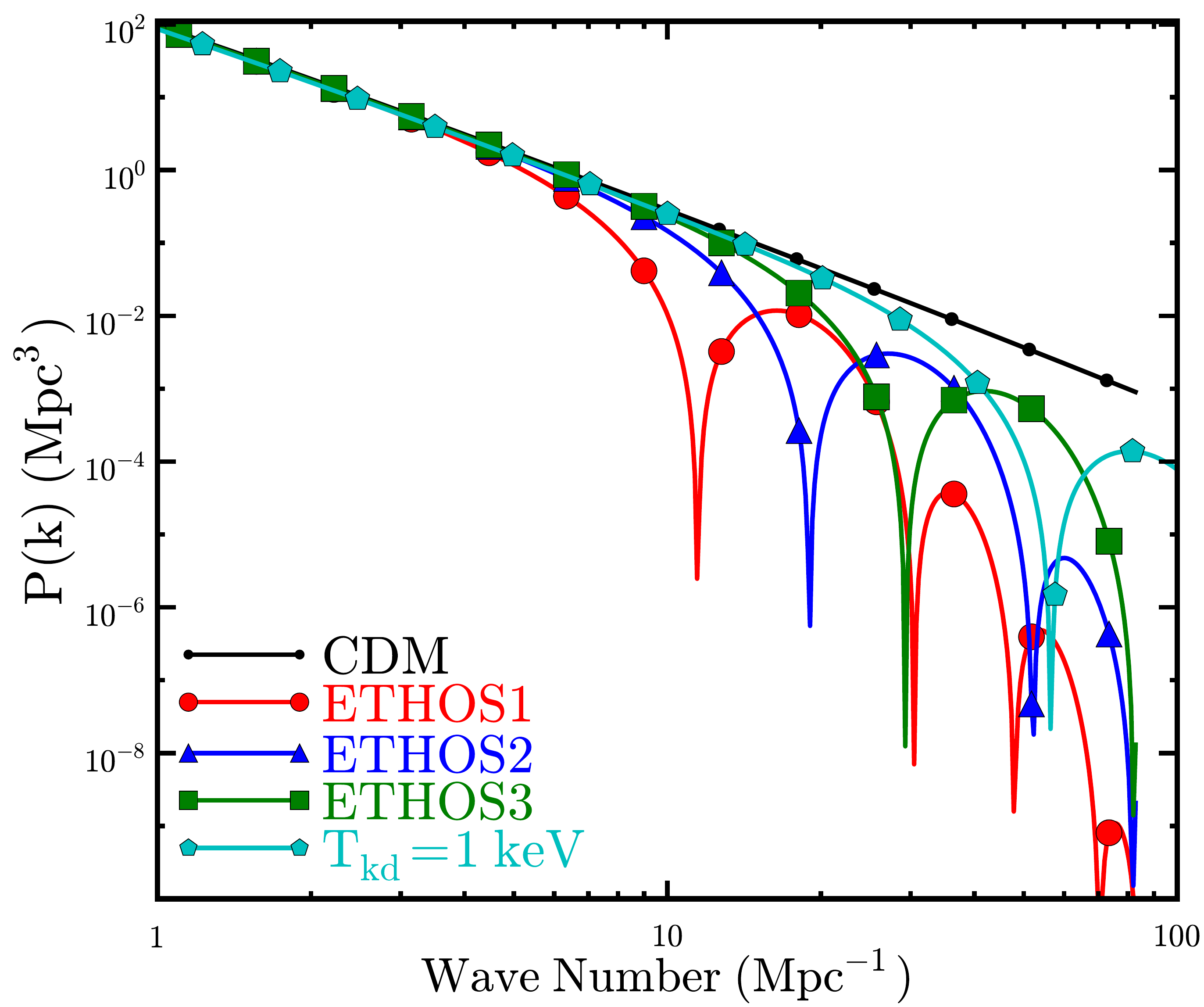}
\caption{Matter power spectra of CDM and interacting DM models, i.e., ETHOS1 ($T_{\rm kd}=0.19~{\rm keV}$), ETHOS2 ($T_{\rm kd}=0.33~{\rm keV}$) and ETHOS3 ($T_{\rm kd}=0.51~{\rm keV}$) from \citetalias{vogelsberger2016}, as well as a model from \citet{huo2017} ($T_{\rm kd}=1~{\rm keV}$).}
\label{fig:MP}
\end{figure}

In Fig.~\ref{fig:MP}, we show the power spectra for the benchmark models in \citetalias{vogelsberger2016}, as well as one in \citet{huo2017}. In these models, DM particles are strongly coupled to relativistic particles (``dark radiation'') in the early universe which results in oscillatory features in their power spectra, analogous to baryonic acoustic oscillations. The damping effect on the power spectra can be characterized by the kinetic decoupling temperature $T_{\rm kd}$ \citep{vandenaarsen2012,cyr-racine2016,huo2017}
\begin{equation}
T_{\rm kd} = \frac{1.38\ {\rm keV}}{\sqrt{g_\chi g_f}}\big(\frac{m_\chi}{100\ {\rm GeV}}\big)^{\frac{1}{4}}\big(\frac{m_\phi}{10\ {\rm MeV}}\big)\big(\frac{g_*}{3.38}\big)^{\frac{1}{8}}\big(\frac{0.5}{\xi}\big)^{\frac{3}{2}},
\end{equation}
at which DM particles kinematically decouple from the radiation plasma ($T_{\rm kd}$ here is defined in terms of the photon temperature, as in \citet{feng2009}). In above equation, $g_\chi$ and $g_f$ are coupling constants, $m_\chi$ and $m_\phi$ are DM and force mediator masses, $g_*$ is the number of massless degrees of freedom at decoupling, and $\xi$ is the ratio of dark-to-visible temperature, $T_f/T_\gamma$. For three ETHOS models, their kinetic decoupling temperatures are $T_{\rm kd}=0.19~{\rm keV}$ (ETHOS1), $0.33~{\rm keV}$ (ETHOS2) and $0.51~{\rm keV}$ (ETHOS3), and the model taken from \citet{huo2017} has $T_{\rm kd}=1~\rm keV$. Fig.~\ref{fig:MP} shows that the suppression on the power spectrum becomes significant as $T_{\rm kd}$ decreases. This is because small $T_{\rm kd}$ indicates a tight coupling between DM and dark radiation, leading to a strong damping effect on the power spectrum. Since the shape and amplitude of DAOs depends on the underlying particle physics of the DM models mostly through the combination that results in the kinetic decoupling temperature, $T_{\rm kd}$ is a viable single parameter to categorize different interacting DM models.

\subsection{Mass variance and window function}\label{sub:mass_variance}
An important ingredient in the Press-Schechter formalism is the mean-squared amplitude of density fluctuations, 
\begin{equation}
\sigma^2(M)\equiv S(M)=\frac{1}{2\pi^2}\int_{0}^{\infty}\mathrm{d}k k^2 P(k) \widetilde{W}^2(k),
\label{eq:sigma}
\end{equation}
where $\widetilde{W}(k)$ is the Fourier transform of the window function. In general, we need to fix $\widetilde{W}(k)$ by comparing the model predictions with N-body simulations. The top-hat filer is a commonly used window function that can successfully reproduce the halo mass function for CDM, see, e.g., \cite{tinker2008,despali2016}. It does, however, produce spurious haloes for DM models with a suppressed power spectrum such as warm DM \citep{benson2013}. An alternative is the sharp-$k$ filter
\begin{equation}
  \widetilde{W}_{\rm sharp\textup{-}k}(k)=
    \begin{cases}
      1 & \text{if}\ k\leq k_{\rm s}(M)\\
      0 & \text{if}\ k> k_{\rm s}(M)\\
    \end{cases},
\end{equation}
where $k_{\rm s}=c/R_0$ with $R_0 \equiv(3M/4\pi\rho_{\rm mean})^{1/3}$. The free parameter $c$ can be fixed by comparing with simulations. However, the sharp-$k$ filter fails to reproduce the halo abundance for the interacting DM models we consider, since it neglects contributions of the modes larger than $k_{\rm s}$, as we will discuss in the next section and Appendix~\ref{sec:massVariance}.
\citet{leo2018} proposed another window function, named as the smooth filter,
\begin{equation}
\widetilde{W}_{\rm smooth}(k)=\frac{1}{1+(k/k_{\rm s})^\beta},
\label{eq:smooth}
\end{equation}
which has two free parameters $\beta$ and $c$ (implicit in $k_{\rm s}$) to be fixed. For large modes (small $k$), it approaches to $1$, similar to the top-hat and sharp-$k$ filters. While for small modes (large $k$), it has non-zero values. Thus, it takes into account large $k$ modes, which are absent in the sharp-$k$ case. On the other hand, we can eliminate spurious haloes by choosing comparably large value of $\beta$, avoiding the shortcomings of the top-hat filter. In this work, we will take the smooth filter for our main results.

\begin{figure}
\includegraphics[width=\columnwidth]{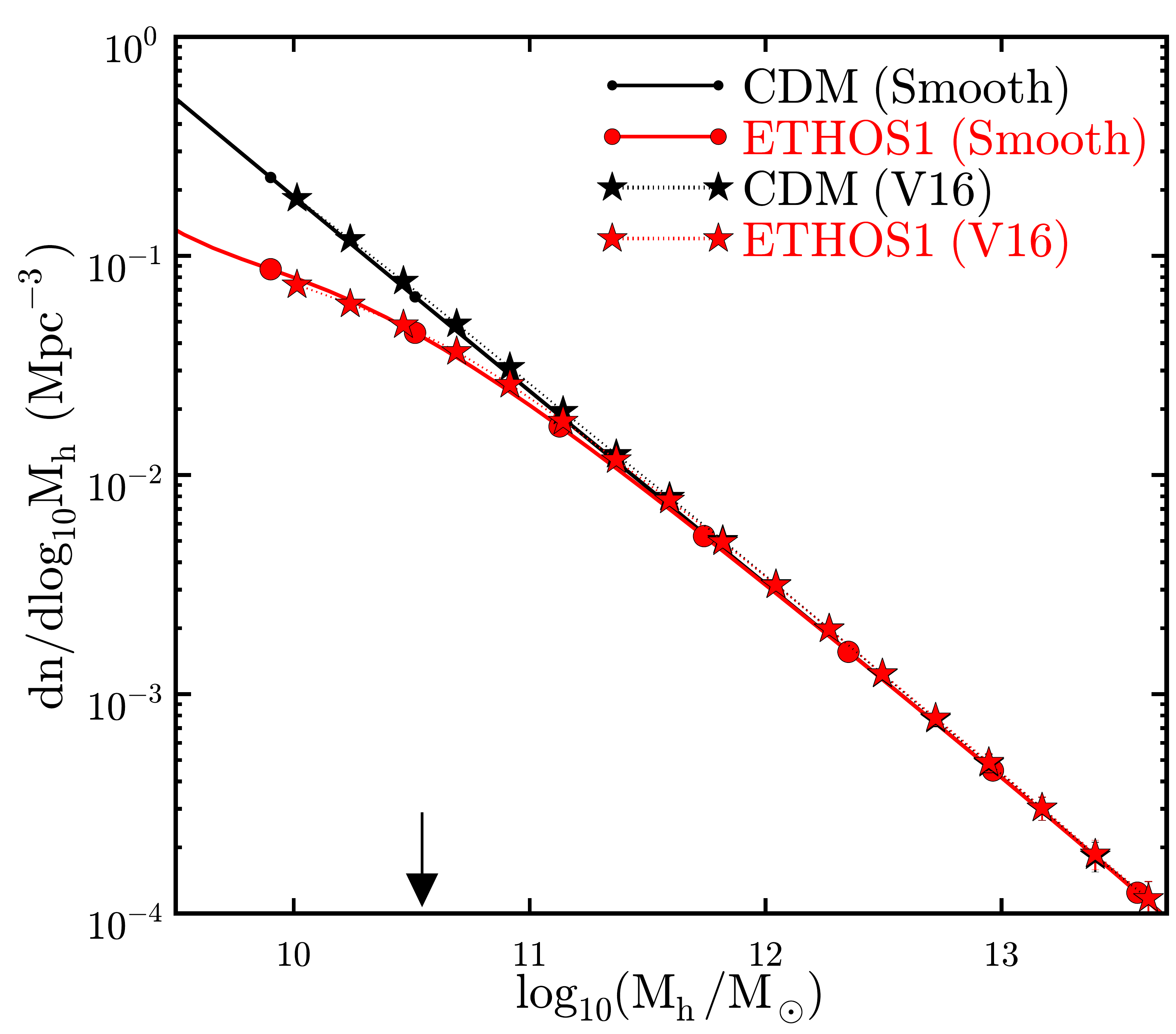}
\includegraphics[width=\columnwidth]{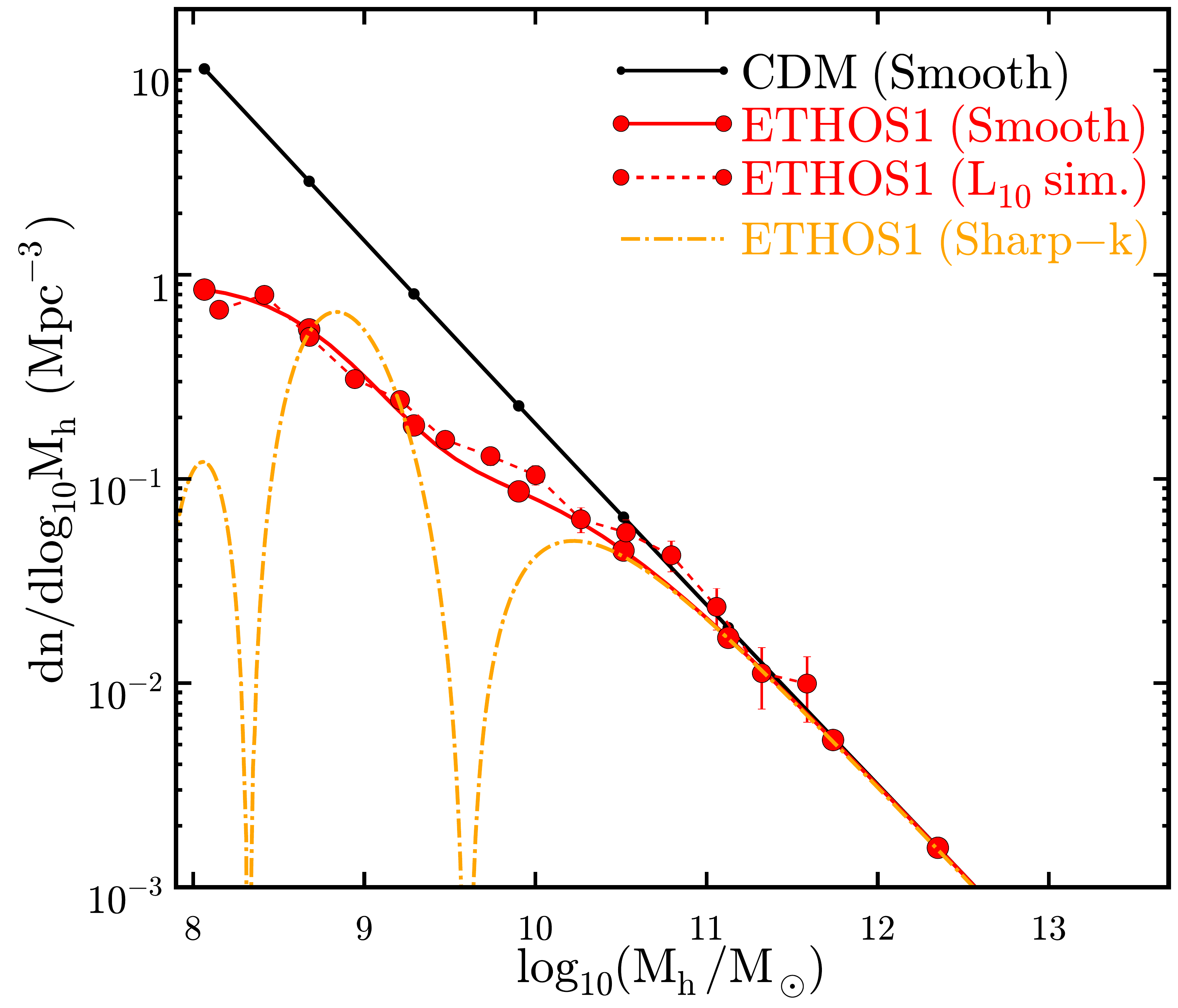}
\caption{
Top: Halo mass functions from the calibrated Press-Schechter model with the smooth filter (solid) and the simulations (dashed) in \citetalias{vogelsberger2016} at $z=0$. Error bars are estimated by assuming that the halo abundance in each mass bin follows Poisson statistics. The black arrow shows $300$ times particle mass $m_{\rm p}$ for simulations in\citetalias{vogelsberger2016}, the lower limit of the mass range we considered for our calibration analysis. Bottom: halo mass functions from the analytical model for ETHOS1 with the smooth (red solid) and sharp-$k$ filters (orange dot-dashed) vs. ${\rm L}_{10}$ simulations carried out in this work (red circles). Our analytical predictions agree with the simulations for the smooth filter. While, the sharp-$k$ filter fails to reproduce the simulation results in the low-mass regimes.
}
\label{fig:best_fit}
\end{figure}
 
Lastly, we comment on the filter-independent approach proposed in \citet{chan2017}. It constructs the shape of the window function using the density profiles of overdense regions, destined to collapse to halos, in the initial linear density field. While this approach provides an independent way to directly measure the shape of the filter, we find that for the interacting DM models it tends to produce spurious haloes. This is because the reconstructed effective filter is essentially the top-hat filter but with edges smoothed by a Gaussian profile and it over predicts the halo abundance for DM models with a cutoff in their power spectra. 
\section{Results}\label{sec:results}
\subsection{Calibrating the model for mass functions}\label{sec:HMF}
We employ the fitting formula for the distribution of first crossing \citep{sheth1999,smt2001} 
\begin{equation}
\nu f(\nu)= 2 A_0 \Big(1+\frac{1}{\nu'^{2p}}\Big)\bigg(\frac{\nu'^2}{2\pi}\bigg)^{1/2} {\rm e}^{-\frac{\nu'^2}{2}},
\label{eq:nufnu}
\end{equation}
with $\nu'=\sqrt{a}\ \nu$, $\nu=\delta_c/\sigma(M)$, and $\delta_{\rm c}=1.686$, to compute the number density of collapsed overdensity peaks per logarithmic mass bins
\begin{equation}
\frac{\mathrm{d}n}{\mathrm{d}\ln{(M)}}=\frac{1}{2}\frac{\rho_{\rm mean}}{M}\nu f(\nu)\frac{\mathrm{d}\log{(\nu)}}{\mathrm{d}\log{(M)}}.
\label{eq:dndlogM}
\end{equation}
We determine the parameters $(A_0,a,p)$ along with $c$ and $\beta$ in Eq.~\ref{eq:smooth} by calibrating the analytical predictions to the N-body simulations. To fix $(c,\beta)$ completely, it is necessary to {\it simultaneously} include both CDM and interacting DM halo abundances in the calibration analysis.

We compute analytical halo mass functions and perform a $\chi^2$ analysis using the joint data points of the CDM and ETHOS1 mass functions from simulations in \citetalias{vogelsberger2016}. We assume that the abundance of the simulated haloes in each mass bin follows a Poisson distribution and, most importantly, we only include in the calibration those mass bins larger than $300 m_{\rm p}\simeq3.34\times10^{10}\; \rm M_{\odot}$, where $m_{\rm p}\simeq1.13\times10^8\; \rm M_{\odot}$ is the particle mass in the ETHOS simulations. We also require the mass bins to contain at least $30$ haloes to minimize the noise from cosmic variance at the high-mass end. Our best-fit values are $(A_0,a,p)=(0.3,0.81,0.3)$ and $(c,\beta)=(3.7,3.5)$ for the smooth filter, see Fig.~\ref{fig:best_fit} (top) for the comparison. We emphasize that although the fit is performed for haloes more massive than $M_{\rm h} \sim 3 \times 10^{10}\; \rm M_{\odot}$ in CDM and ETHOS1 simulations at $z=0$, we will use the model to {\it predict} the abundances of lower mass haloes as well, for different ETHOS models at different redshifts and for wider range of halo masses.

To better understand the effects of different filters on the predicted halo mass functions, we compare our simulated results ($L_{10}$) with analytical predictions for the smooth and sharp-$k$ filters given a wider range of halos masses, as shown in Fig.~\ref{fig:best_fit} (bottom). Although the analytical mass function with the sharp-$k$ filter ($c=3.2$) is in reasonable agreement with the simulated one at the high-mass end, it exhibits significant oscillatory features and fails in low-mass halo regimes. Indeed, for the sharp-$k$ filter, $\mathrm{d}n(M)/\mathrm{d}\log_{10}{(M)}\propto P(k_{\rm s})$, i.e., the shape of mass function closely follows the power spectrum \citep[see][for similar results]{schewtschenko2015,leo2018}. On the other hand, the smooth filter smoothes out the dark acoustic peaks and the result agrees with the simulations remarkably well. It has been shown that non-linear evolution of the modes erases the peaks in the power spectrum predicted in interacting DM models \citep{buckley2014}, resulting in a smooth decay in halo number density. It seems the smooth filter accurately captures this effect by including contributions of all modes to the mass variance, as we have shown. The Press-Schechter method with the smooth filter also successfully captures the depletion of halos on scales well below the halo mass associated with the first trough. We have checked that other analytical models \citep[e.g. see][]{vogelsberger2016}, where a simple exponential decay exp (−Mcut/M) is applied to the CDM mass function, fail to reproduce halo abundances for mass scales lower than the first trough.   

\begin{figure*}
\includegraphics[width=1.01\columnwidth]{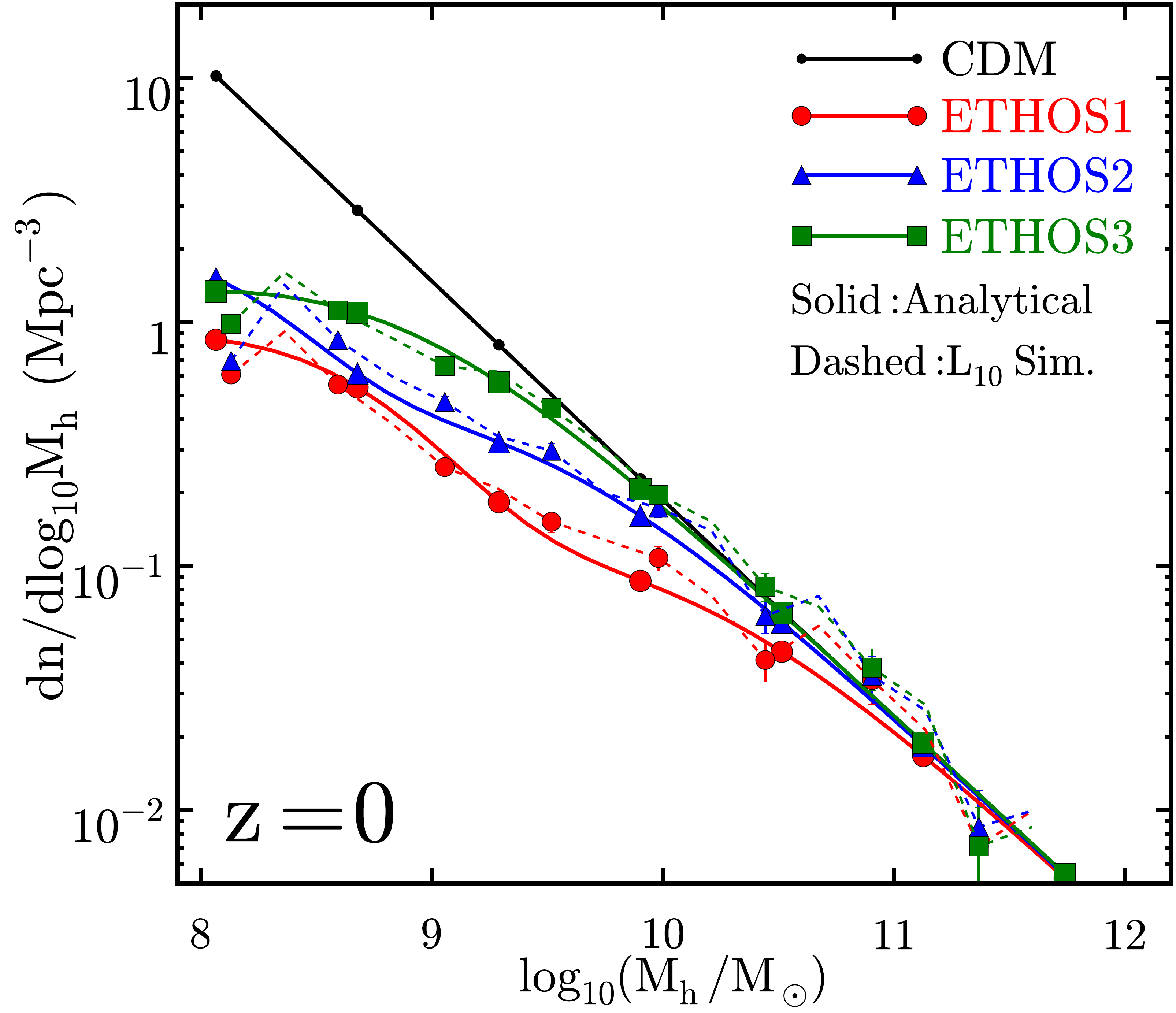}
\includegraphics[width=1.01\columnwidth]{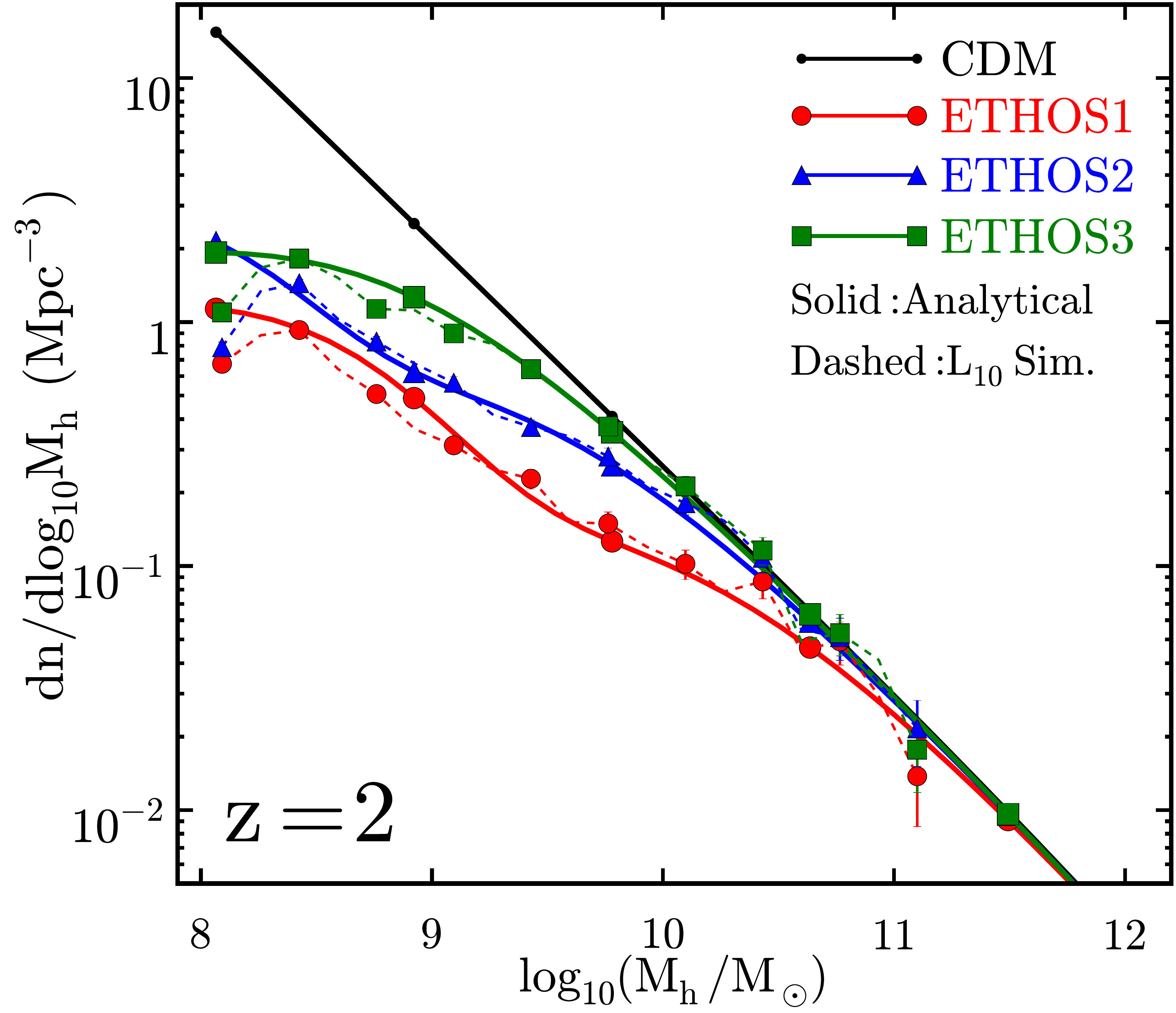}\\
\includegraphics[width=1.01\columnwidth]{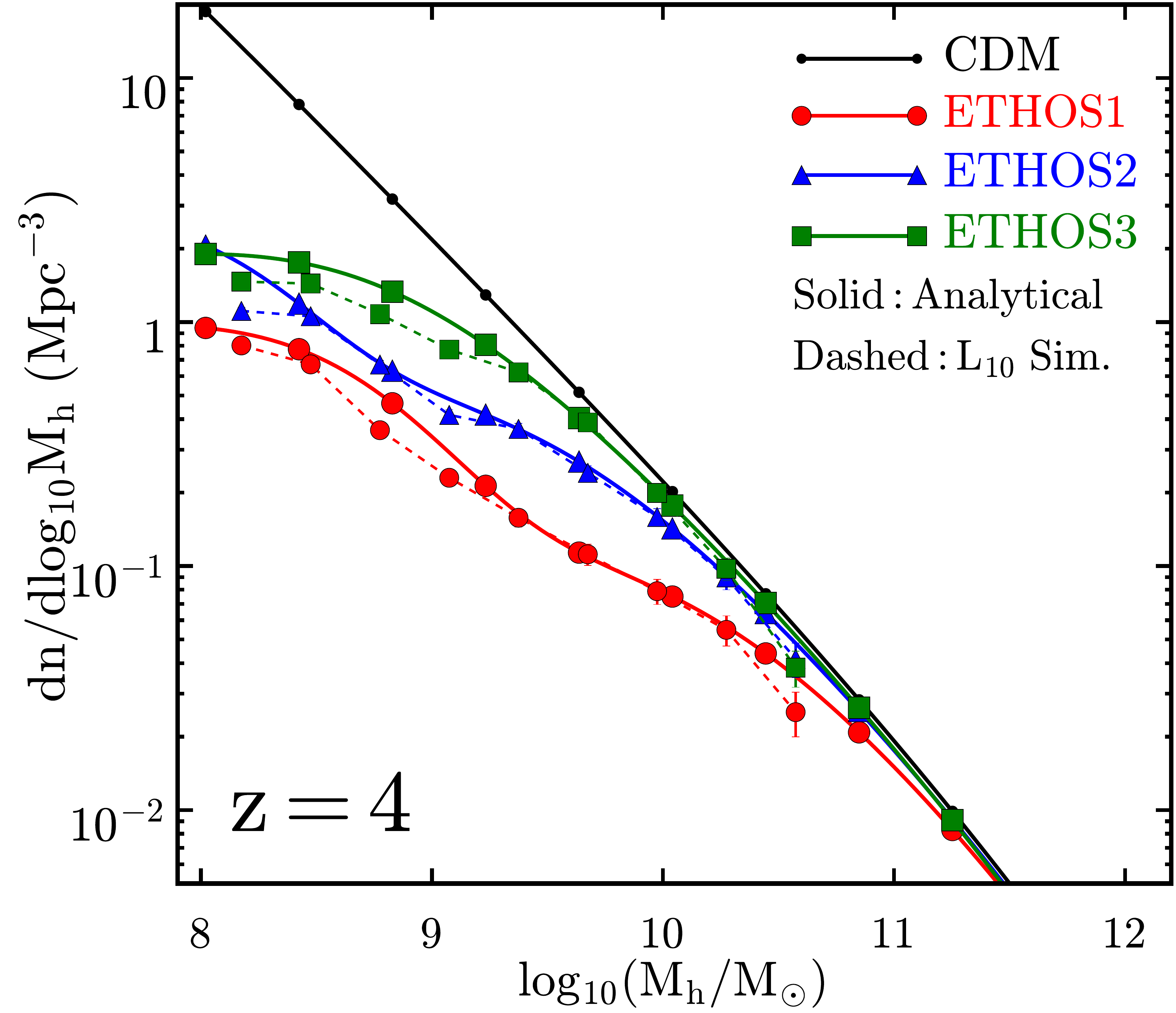}
\includegraphics[width=1.01\columnwidth]{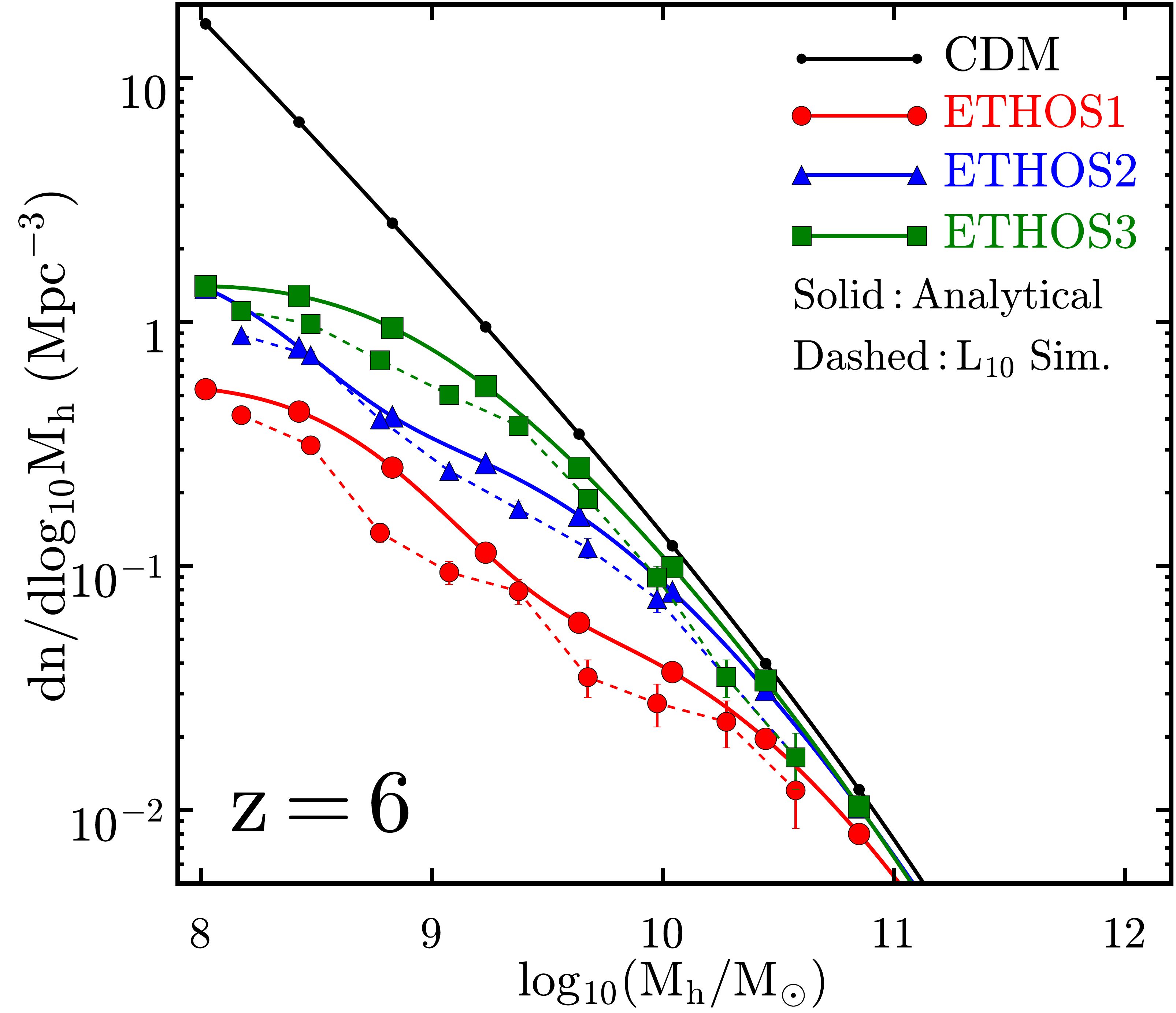}
\caption{
Analytical halo mass functions (solid) vs. simulation results (dashed) at four different redshifts $z=0$, $2$, $4$ and $6$.  The analytical model, after calibrated with the simulated mass functions of CDM and ETHOS1 down to $3\times10^{10}\;\rm M_{\odot}$ at $z=0$, can successfully reproduce the simulations with halo masses down to $10^8\; \rm M_{\odot}$ at different redshifts for two other interacting DM models, ETHOS2 and ETHOS3. It slightly over predicts the halo abundances at $z=6$. }
\label{fig:high_z}
\end{figure*}

Fig.~\ref{fig:high_z} shows excellent agreement between our analytical predictions and ${\rm L}_{10}$ simulations introduced in Sec.~\ref{sec:method} at different redshifts. We emphasize again that our model parameters are constrained to reproduce ETHOS1 and CDM simulations at $z=0$ for halo masses larger than $\sim3 \times 10^{10} \; \rm M_\odot$, and the higher redshift and lower-mass comparison demonstrate how well this single model extends to these regimes. The analytical model slightly over predicts the number density of haloes at $z=6$. This $20\textup{--}30\%$ discrepancy toward high redshifts has been noted in other works \cite[e.g.][]{courtin2011,despali2016}. It is evident that the halo abundance is more suppressed for DM models with a lower kinetic decoupling temperature, as expected.

Our model will allow us to make predictions for halo abundances at low and high redshifts and for different cosmology assumptions. This is particularly interesting in light of the availability of the Wide Field Camera 3 (WFC3) on the Hubble Space Telescope (HST) combined with gravitational lensing effects from clusters in the Hubble Frontier Fields, which have provided exquisite measurements of the galaxy luminosity function on the UV down to magnitudes of $\rm M_{\rm UV} \simeq -15$ \citep[e.g. see][]{mclure2013,bouwens2015,finkelstein2015}, comparable to what is possible only within the Local Volume. In what follows, we take advantage of these high-$z$, volume-complete observational constraints for faint galaxies and use our analytical approach to compare with the abundance of low mass haloes predicted for different DM models. 

\begin{figure}
\includegraphics[width=\columnwidth]{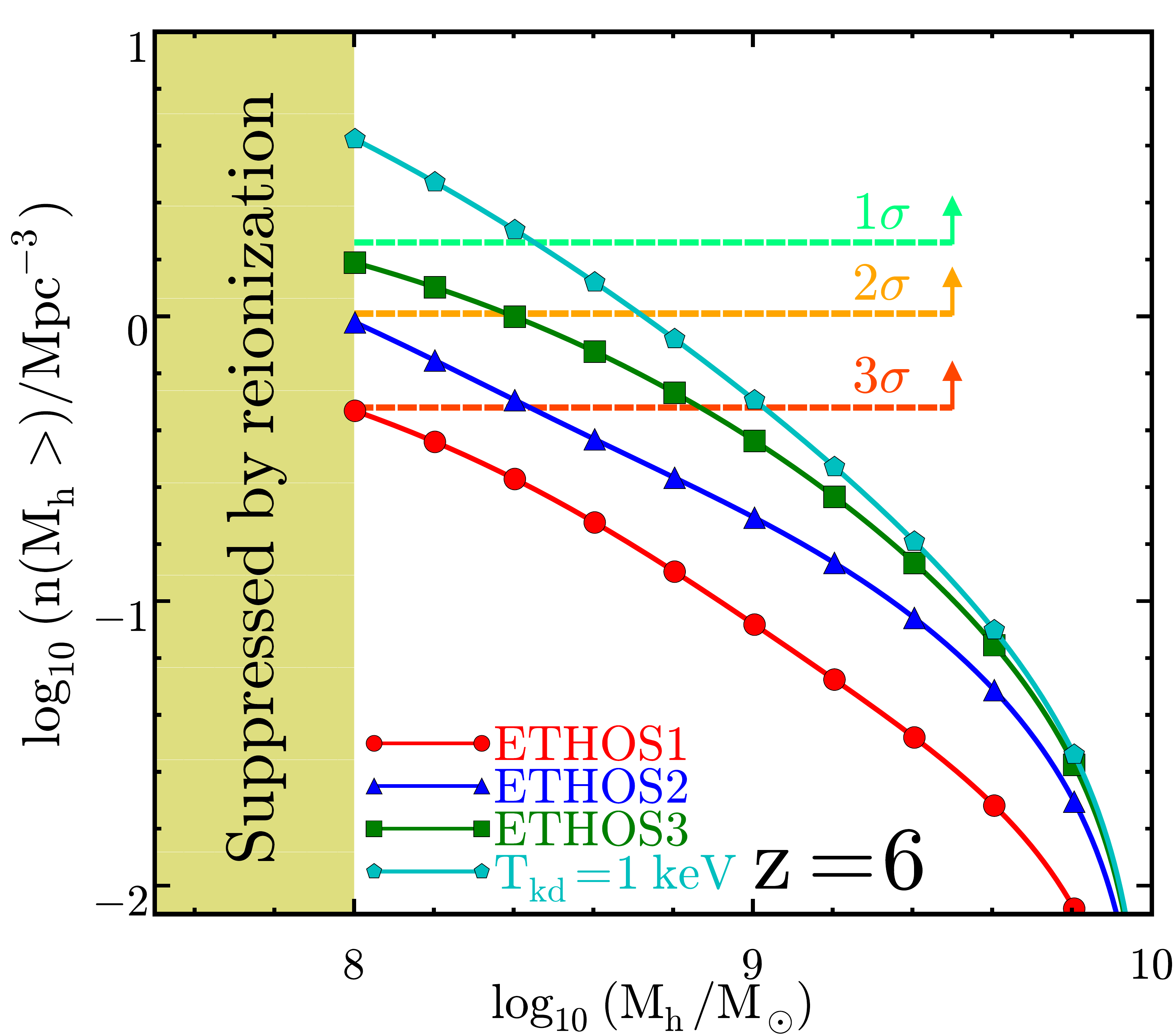}
\caption{
Cumulative number density of haloes, $n(>M)$, for different interacting DM models vs. cumulative galaxy number density estimated from observed UV luminosity function $n_{\rm obs}(>M)$ at $z=6$ \citep{menci2016b}, where the horizontal lines denote the confidence levels of the lower bound on the galaxy counts. The shaded region denotes the halo mass scales, i.e. $M_{\rm h}\leq 10^8\; \rm M_{\odot}$, which have lost most of their baryons due to background UV radiation \citep{okamoto2008,ocvirk2016}.
}
\label{fig:obs_constraint}
\end{figure}

\begin{figure}
\includegraphics[width=\columnwidth]{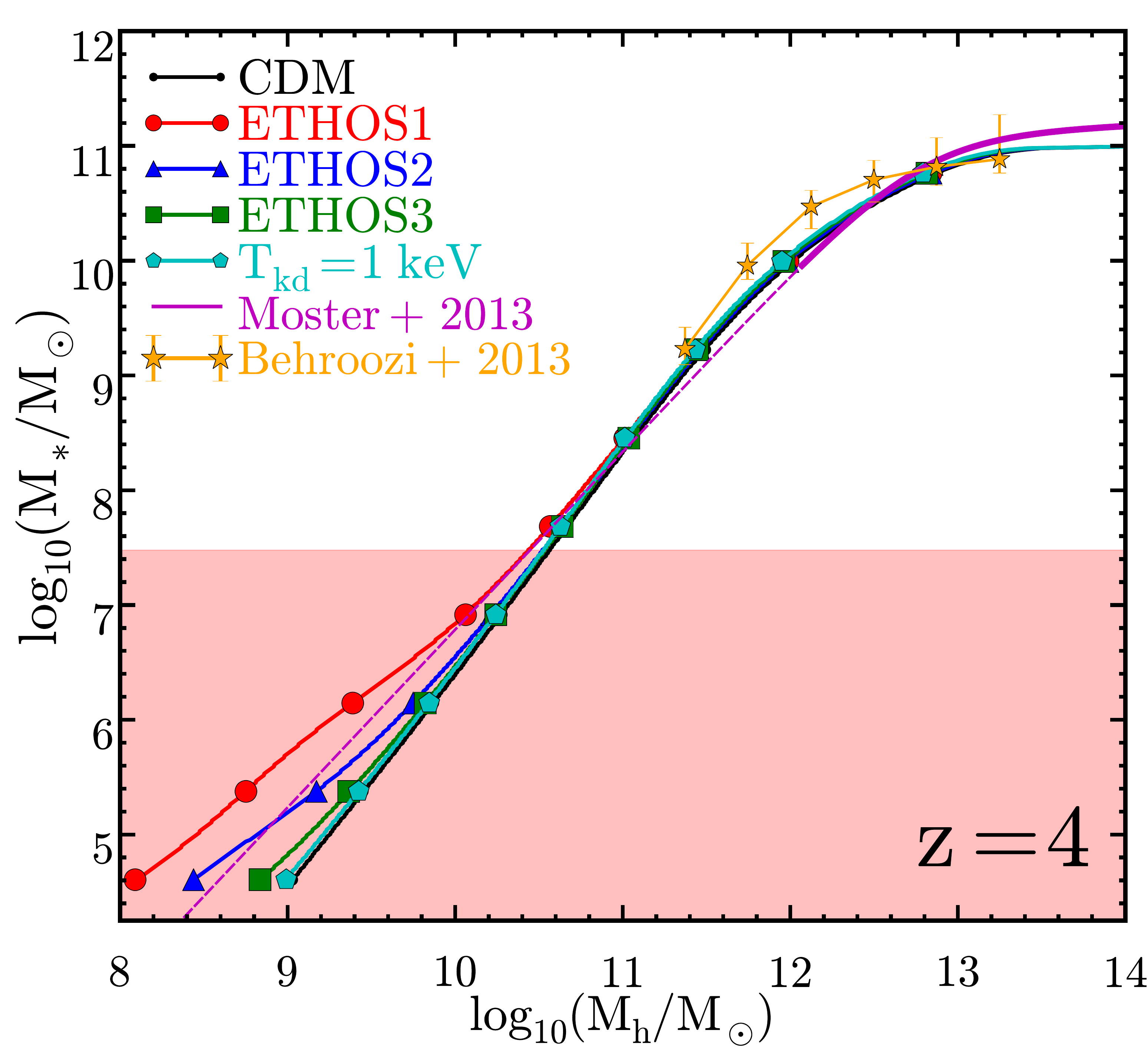}
\caption{Predicted stellar-halo mass relations for CDM, ETHOS models and the DM model with $\rm{T_{kd}=1\ keV}$ at $z=4$. We also show the results for CDM in \citet{behroozi2013} and \citet{moster2013} (magenta dashed: extrapolated). The shaded region indicates the mass scales with the extrapolated stellar mass function.
}
\label{fig:stellar_halo}
\end{figure}

\subsection{Constraining the DM models with galaxy abundance at $z=6$}
\label{sub:uv_lum}

In Fig.~\ref{fig:obs_constraint}, we show the cumulative number density of haloes $n(>M)$ derived from our analytical model, together with the constraints on the number density of galaxies from \citet{menci2016b}, which is based on the luminosity functions in \cite{livermore2017}. The horizontal lines denote the confidence levels of the observational constraints. In the shaded region, where the mass is below $10^8\; \rm M_{\odot}$, we expect DM haloes to have lost most of their baryons due to photoheating caused by the ionizing background UV radiation at $z=6$ \citep[see, e.g.,][]{okamoto2008,ocvirk2016}. We see that ETHOS1 and ETHOS2 are outside of the $3\sigma$ and $2\sigma$ limits, respectively, and ETHOS3 is marginally consistent within the $1\sigma$ limit, Whilst the model with $T_{\rm kd}=1~{\rm keV}$ is fully within the observational constraints. Interestingly, these lower bounds on $T_{\rm kd}$ from the galaxy counts are coincident with those from the Lyman-$\alpha$ forest observations \citep{huo2017}. 

\citet{lovell2017} find the UV luminosity function to be indistinguishable between CDM and ETHOS4 down to $M_{\rm UV}\leq-13$, or $M_*\geq 3\times 10^5\ M_{\odot}$ (by taking the median of the $M_{\rm UV}-M_*$ relation from \citet{song2016} and extrapolating it to $M_{\rm UV}=-13$) at $z=6$. These stellar masses translate to halo masses of $M_{\rm h}\geq10^9 M_{\odot}$ for this redshift \citep{song2016}. This is consistent with our results in Fig.~\ref{fig:obs_constraint}, which suggest that halo masses $\sim 10^8\ M_\odot$ should be reached in order to observationally rule out some of the most extreme non-CDM models such as the ETHOS1.

\subsection{Stellar mass-halo mass relation}\label{sub:smhmr}

As another application, we derive the stellar-halo mass relation for the interacting DM models using the abundance matching technique, i.e., matching cumulative halo plus subhalo mass functions to the observed number density of galaxies in each stellar mass bin,
\begin{equation}
n(>M_*)= n_{\rm h}(>M)+n_{\rm sh}(>M).
\end{equation} 
Our analytical model does not include the presence of subhaloes deemed subdominant but important for these kinds of calculations. We therefore aim at estimating their effects as follows. The CDM subhalo mass function in each mass is calculated as
\begin{equation}
n_{\rm sh;CDM}(m,z)=\int_{0}^{\infty}N(m|M,z)n_{\rm h}(M,z) \mathrm{d}M,
\end{equation}
where $N(m|M,z)\mathrm{d}m=N(m|M,z=0)f(z) \mathrm{d}m$ is the total number of subhaloes in the mass range $m+\mathrm{d}m$ for a parent halo with mass $M$ at given redshift $z$. We take the analytical formula for the subhalo distribution for a given parent halo at $z=0$ from \citet{giocoli2008},
\begin{equation}
N(m|M,z=0) = \frac{N_0}{m} x^{-\alpha}{\rm e}^{-6.283 x^3}, x=\frac{m}{\alpha M},
\label{eq:subdis}
\end{equation}
where $\alpha=0.8$ and $N_0=0.21$. To extend this to earlier times we normalise Eq.~\ref{eq:subdis} by the redshift evolution factor $f(z)\equiv f_{\rm sub}(z)/f_{\rm sub}(z=0)=1-z/6$ \citep{conroy2009} and neglect mild dependence of $f(z)$ on the halo mass. Note that the subhalo distribution function given in Eq.~\ref{eq:subdis} is calibrated with CDM simulations. In principle, one needs to recalibrate it for the interacting DM models as well. For simplicity, we assume the depletion rate of subhaloes in the non-CDM models is as the same as that of main halos, and estimate the subhalo distribution as\footnote{ This estimation, albeit oversimplified, is equivalent to assuming that the fraction of subhalos to main halos at a given mass is the same as CDM, a claim that shall be confirmed with simulations in the future. However, we do not expect the additional refinement to the current approach will change the main conclusion in this work since the contribution of subhalos for these redshifts is at the level of a few percent \citep{conroy2006}.}
\begin{equation}
n_{\rm sh;non\textup{-}CDM}(m,z) = \frac{n_{\rm h;non\textup{-}CDM}(M,z)}{n_{\rm h;CDM}(M,z)}n_{\rm sh,CDM}(m,z).
\end{equation}  
We also neglect the effect on the subhalo mass function caused by the finite resolution of simulations \citep[]{guo2014}. Since subhaloes have sub-dominant contributions to the total mass function, we expect that our estimate of $n_{\rm sh;non\textup{-}CDM}(m,z)$ will provide a reasonable approximation.

For the number density of observed galaxies, we take the Schechter \citep{schechter1976} function
\begin{equation}
\frac{\mathrm{d}n(M_*)}{\mathrm{d}\log_{10}{(M_*)}}=\ln{(10)}\phi^*{\rm e}^{-\frac{M_*}{M_0}}\Big(\frac{M_*}{M_0}\Big)^{\alpha+1}, 
\label{eq:schechter_fit}
\end{equation}
where $(\log_{10}{(M_0)},\alpha,\log_{10}{(\phi^*)})=(10.5,-1.55,-3.59)$ are the best-ft values after fitting to the stellar mass function for their sample of galaxies at $z\sim 4$ \citep{song2016}.

In Fig.~\ref{fig:stellar_halo}, we show the stellar-halo mass relations for the DM models we consider after matching the number density of galaxies to that of haloes. For comparison, we also plot the CDM results from \citet{behroozi2013} and \citet{moster2013} (magenta solid: stellar-halo mass relation in the mass range where observational data points exist; magenta dashed: extrapolation to lower masses). Overall, our result for CDM is in reasonable agreement with previous works in the mass region spanned by the observational data points. The small offset could be caused by different choices of galaxy samples, halo mass definitions and window functions. We have checked that using a top-hat filter slightly improves the agreement for the CDM case.
 
The suppression of the halo mass function results in more massive galaxies inhabiting a given DM halo or, in other words, a higher star formation efficiency in the low mass end. This can be seen clearly for the interacting DM models in Fig.~\ref{fig:stellar_halo}, with ETHOS1 being the most strongly suppressed and therefore exhibiting larger deviations from the CDM results. The red shaded region shows the regime where the observed luminosity function has been extrapolated using the Schechter function. Our results indicate that significant deviations from the other models in the case of ETHOS1 could be achieved by reaching observational completeness in the range $M_{*}\sim 10^6\textup{-}10^7\; \rm M_{\odot}$, about a dex fainter than current limits. On the other hand, the limits are fainter for ETHOS2 ($M_{*}\sim 10^5\; \rm M_{\odot}$), while ETHOS3 and the DM model with $T_{\rm kd}=1~{\rm keV}$ seem indistinguishable from CDM stellar-halo mass relation down to very faint limits of $M_{*}\sim 10^4\; \rm M_{\odot}$. The effort, coupled with an alternative measurement of halo masses, such as clustering and kinematics, could be used to further test DM models with suppressed matter power spectra. And our methodology and prescriptions to fast compute halo mass functions may prove useful for these kinds of assessments.
\\
\\
\section{Summary}\label{sec:conc}
We have used the Press-Schechter formalism to study the halo mass functions for interacting DM models with matter power spectra damped by dark acoustic oscillations. Taking three ETHOS models as benchmark examples, we have demonstrated that this analytical approach can accurately reproduce the result of N-body simulation results. The choice of a proper window function plays a critical role in such as success. We found the smooth filter proposed in~\citet{leo2018} works well in capturing relevant physics. The sharp-$k$ filter, despite being a viable choice for warm DM, fails in the low-mass regimes for interacting DM. Our model parameters are constrained to match the CDM and ETHOS1 simulations at the high mass end of the $z=0$ mass functions. In order to validate these we performed our own cosmological simulations with improved mass resolution. Our results indicate that the Press-Schechter formalism with the smooth filter provides a simple but powerful tool to understand the suppression effect on the halo mass functions induced by DM-dark radiation interactions, presented in many new DM models beyond the CDM paradigm.

We have further applied our calibrated model to derive constraints on the DM models using the observed stellar mass functions at high redshifts available in the literature. After comparing the cumulative number density of haloes predicted in the DM models with that of galaxies inferred from the measured UV luminosity functions at $z=6$ \citep{menci2017,livermore2017}, we found both ETHOS1 ($T_{\rm kd}=0.19~{\rm keV}$) and ETHOS2 ($T_{\rm kd}=0.33~{\rm keV}$) strongly disfavored, as they produce too few halos due to host the observed galaxies, due to strong dark acoustic damping. While ETHOS3 ($T_{\rm kd}=0.51~{\rm keV}$) and the model with $T_{\rm kd}=1~{\rm keV}$ are within the observational constraints. Interestingly, these UV luminosity constraints on the kinetic decoupling temperature of the interacting DM models are similar to those from the Lyman-$\alpha$ forest measurements reported in \cite{huo2017}. 

We have also performed an abundance matching analysis to derive the stellar-halo mass relation for the interacting DM models, using the observed stellar mass functions of galaxies at $z=4$ \citep{song2016}. Our results indicate appreciable suppression in the halo mass of hosted galaxies at $M_{*}\lesssim10^5\textup{-}10^7\; \rm M_{\odot}$ for ETHOS1 and ETHOS2 models, and ETHOS3 shows mild suppression of halo mass only in the low-mass tail $M_{*}\lesssim10^5\;\rm M_{\odot}$. In contrast, the DM model with $T_{\rm kd}=1\ \rm{keV}$ is almost indistinguishable from CDM. While it is of great interest to further push  observational limits to dwarf galaxies below $M_*\sim 10^5\ {\rm M}_\odot$, our model provides a valuable tool to explore interacting DM models quickly and with low computational demands; facilitating the comparison between observations and theoretical expectations in the quest to determine the nature of dark matter.

\section*{Acknowledgments}
We thank Volker Springel for making {\sc Arepo} available for this work, and Mark Lovell for providing mass functions from the ETHOS project for comparison. We also thank Mark Vogelsberger, Jes\`us Zavala, Francis-Yan Cyr-Racine, and Simeon Bird for useful comments and discussion. OS acknowledges support by NASA MUREP Institutional Research Opportunity (MIRO) grant number NNX15AP99A and HST grant HST-ART-14582. LVS is grateful for support from the Hellman Fellows Foundation and HST grant HST-ART-14582. HBY acknowledges support from U.S. Department of Energy under Grant No. de-sc0008541 and UCR Regents' Faculty Development Award. The work of LAM was carried out at Jet Propulsion Laboratory, California Institute of Technology, under a contract with NASA.
\begin{appendix}
\medskip
\begin{figure}
\includegraphics[width=\columnwidth]{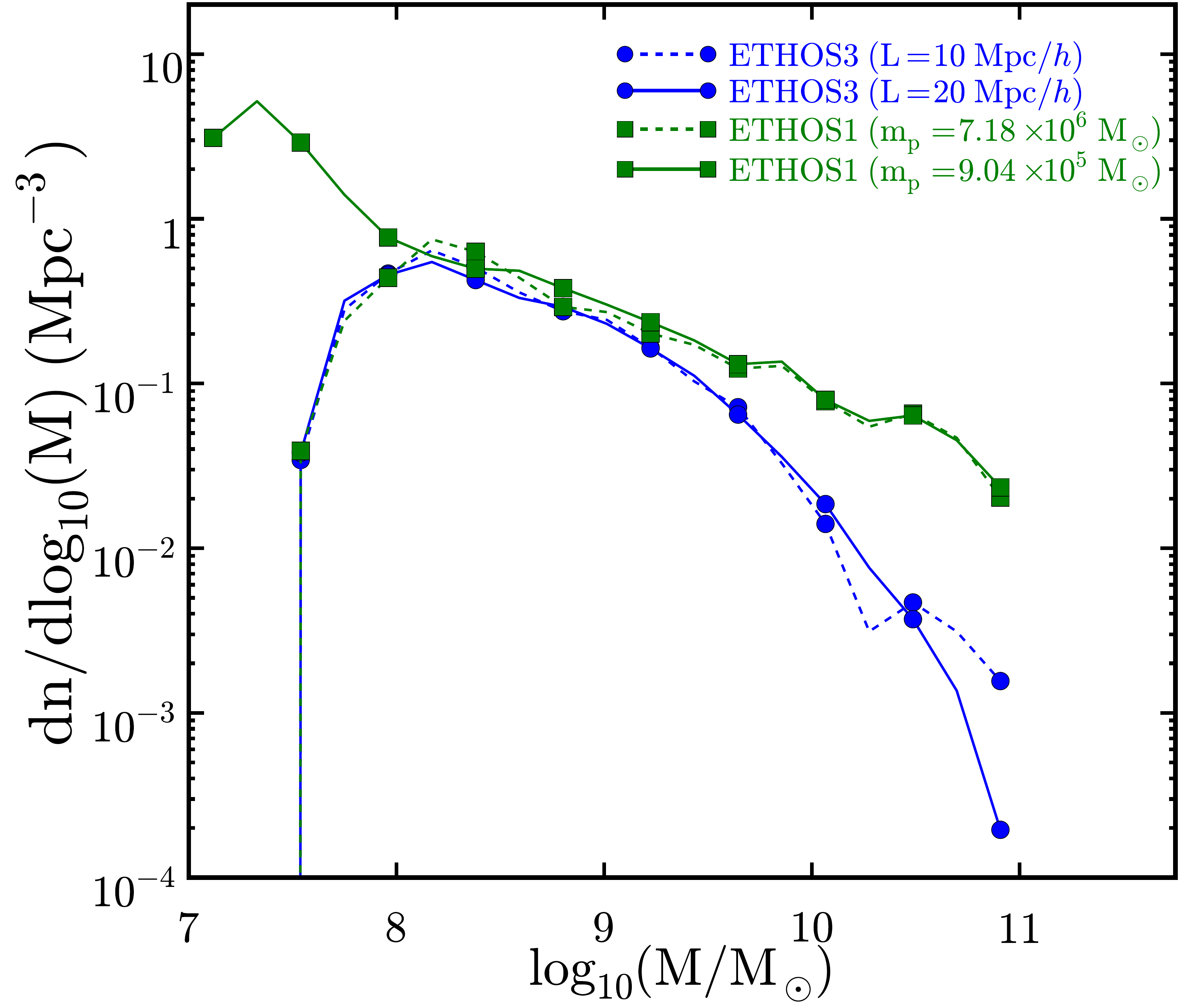}
\caption{
Convergence test of our cosmological simulations for two different DM models at different redshifts. Green lines denote the halo mass functions for ETHOS1 with box size ${\rm L}=10\ {\rm Mpc}/h$ and particle mass resolution $m_{\rm p}=7.18\times 10^6\ {\rm M}_{\odot}$ (dashed) and $9.04\times 10^5\; \rm M_{\odot}$ (solid) at $z=0.2$. Blue lines denote ETHOS3 simulations with box sizes ${\rm L}=10~{\rm Mpc}/h$ (dashed) and $20 ~{\rm Mpc}/h$ (solid) and the same mass resolution $m_{\rm p}= 7.18\times 10^6\; \rm M_{\odot}$ at $z=8$.
}
\label{fig:resolution}
\end{figure}
\begin{figure}
\includegraphics[width=\columnwidth]{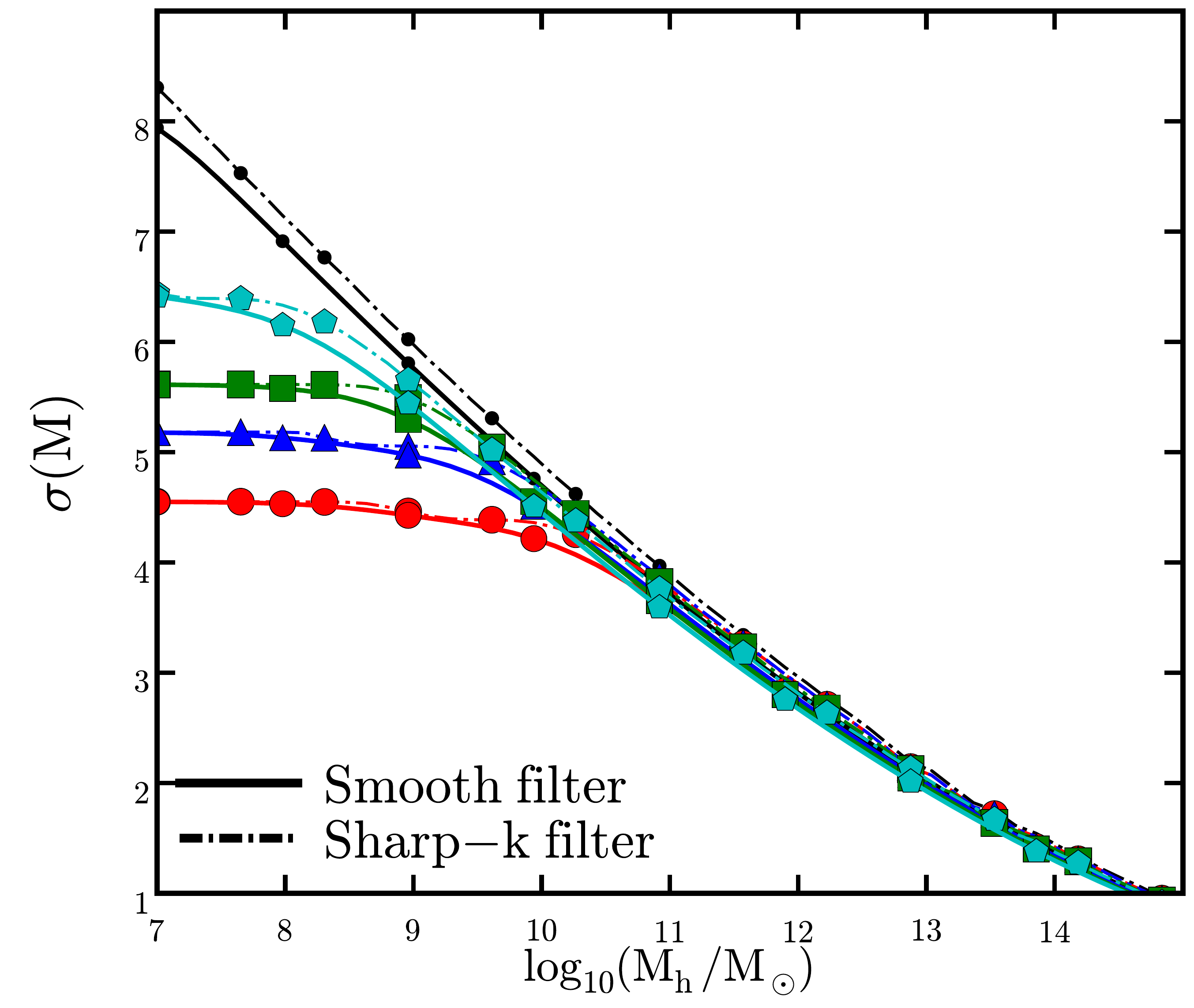}
\includegraphics[width=\columnwidth]{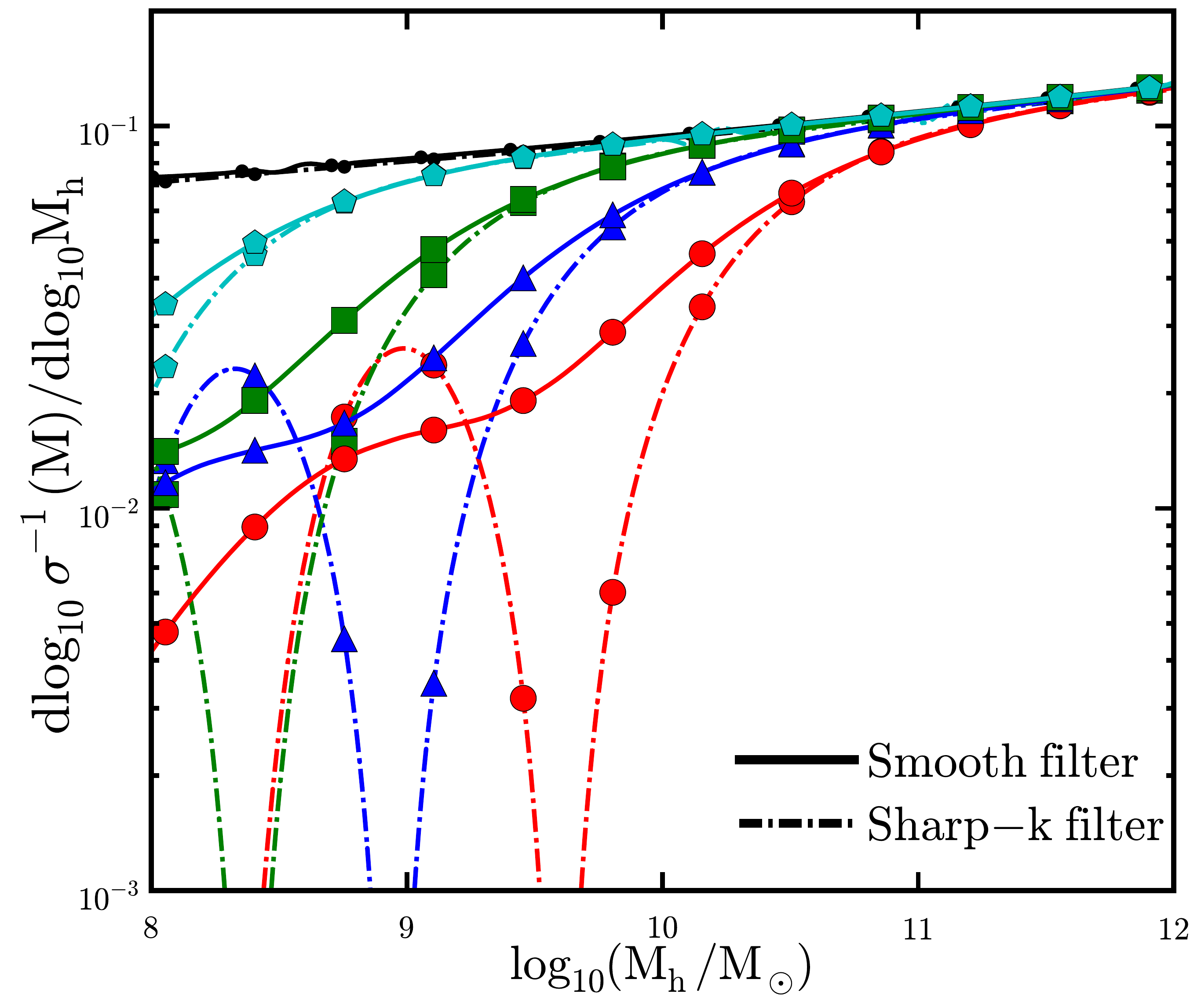}
\caption{Top: Mass variance calculated for the interacting DM models with the sharp-$k$ (dot-dashed) and smooth filters (solid). Bottom: $\mathrm{d}\log_{10}{\sigma^{-1}(M)/\mathrm{d}\log_{10}M}$ for the DM models. Marker styles and color schemes are the same as in Fig.~\ref{fig:MP}.
}
\label{fig:massVariance}
\end{figure}
\section{Convergence and resolution test}
\label{sec:A1}
We test the numerical convergence of our simulations. In Fig.~\ref{fig:resolution}, we show simulated halo mass functions for ETHOS1 with two different levels of mass resolution: $m_{\rm p}=7.18\times 10^6\; \rm M_{\odot}$ (green dashed) and $9.04\times10^5\; \rm M_{\odot}$ (green solid) at redshift $z\sim 0.2$. The cosmological box size is ${\rm L}=10\ \rm{Mpc}$. We find good numerical convergence down to $10^8\; \rm M_{\odot}$. For lower halo masses, the low resolution simulation cannot populate halos, while the high resolution one suffers from spurious haloes. The slight overabundance of halos in the low resolution simulation for ETHOS1 in some of the low-mass bins is likely due to statistical noise. We have checked that the mass functions for low and high resolution simulations are consistent within $1\textup{--}3\sigma$ assuming Poisson error bars. Moreover, if we increase the size of mass bins (thereby increasing the signal-to-noise in each bin), the difference becomes even smaller.

We also test the effect of cosmic variance by comparing the halo mass function for two ETHOS3 simulations with the same mass resolution ($m_{\rm p}= 7.18\times10^6\; \rm M_{\odot}$) but with different box sizes ${\rm L}=10\ {\rm Mpc}/h$ (blue dashed) and $20\ {\rm Mpc/h}$ (blue solid) at $z=8$. They are well converged toward the low-mass end, but deviate for high halo masses because the simulations with a small box size suffers from cosmic variance. We find good convergence of the mass functions in the mass range $10^8\textup{--}10^{10}\; \rm M_{\odot}$, and we take this range in our analysis in Sec.~3.

\section{Mass variance and window function-continued}\label{sec:massVariance}
In Fig.~\ref{fig:massVariance} (top), we show the mass variance for the DM models considered in this work for both the sharp-$k$ space ($c=3.2$; dot-dashed) and smooth filters ($c=3.7\ {\rm and}\ \beta=3.5$; solid). As we have discussed in the Sec.~\ref{sec:HMF}, the prediction of the analytical model in the low-mass regime is sensitive to the choice of the filter. To demonstrate the origin of this effect, we compute $\mathrm{d}\log_{10}{\sigma^{-1}(M)}/\mathrm{d}\log_{10}{M}$, the key factor in Eq.~\ref{eq:dndlogM}, for each DM model with both smooth and sharp-$k$ filters, as shown in Fig.~\ref{fig:massVariance} (bottom). When computed with the sharp-$k$ filter, the factor has an oscillatory feature in the low mass end, a reminiscent of the acoustic peaks in the power spectrum. 

\end{appendix}
\bibliographystyle{yahapj}
\bibliography{paper}
\end{document}